\documentclass[11pt,epsf]{article}
\usepackage{comment}
\newcommand {\dfn} {\stackrel{\Delta} {=}}

\newcommand {\reals} {{\rm I\!R}}

\newcommand {\bx} {\mbox{\boldmath $x$}}
\newcommand {\by} {\mbox{\boldmath $y$}}

\newcommand{\calA}{{\cal A}}
\newcommand{\calB}{{\cal B}}
\newcommand{\calC}{{\cal C}}

\newcommand{\calE}{{\cal E}}

\newcommand{\calI}{{\cal I}}

\newcommand{\calP}{{\cal P}}

\newcommand{\calR}{{\cal R}}
\newcommand{\calS}{{\cal S}}
\newcommand{\calT}{{\cal T}}
\newcommand{\calU}{{\cal U}}
\newcommand{\calV}{{\cal V}}

\newcommand{\calX}{{\cal X}}

\newcommand{\calZ}{{\cal Z}}

\newcommand {\gx} {\grave{x}}
\newcommand {\tx} {\tilde{x}}
\newcommand {\hx} {\hat{x}}
\newcommand {\hX} {\hat{x}}
\newcommand {\tX} {\tilde{X}}
\newcommand {\gX} {\grave{X}}
\newcommand{\hcalX}{\hat{\cal X}}
\newcommand{\tcalX}{\tilde{\cal X}}

\begin{document}
\thispagestyle{empty}
\title{Successive Refinement for Lossy Compression\\
of Individual Sequences
}
\author{Neri Merhav
}
\date{}
\maketitle

\begin{center}
The Andrew \& Erna Viterbi Faculty of Electrical Engineering\\
Technion - Israel Institute of Technology \\
Technion City, Haifa 32000, ISRAEL \\
E--mail: {\tt merhav@ee.technion.ac.il}\\
\end{center}
\vspace{1.5\baselineskip}
\setlength{\baselineskip}{1.5\baselineskip}

\begin{abstract}
We consider the problem of successive-refinement coding for lossy
compression of individual
sequences, namely, compression in two stages,
where in the first stage, a coarse description at a relatively low rate is
sent from the encoder to the decoder,
and in the second stage, additional coding rate is allocated in order to
refine the description and thereby improve the reproduction. Our main result
is in establishing outer bounds
(converse theorems) for the rate region where
we limit the encoders to be finite-state machines in the spirit of Ziv and
Lempel's 1978 model.
The matching achievability
scheme is conceptually straightforward.
We also consider the more general multiple
description coding problem on a similar footing and propose achievability schemes that are
analogous to the well-known El Gamal-Cover and the Zhang-Berger achievability schemes
of memoryless sources and additive distortion measures.
\end{abstract}

\section{Introduction}
\label{intro}

The notion of {\em successive refinement} of information refers to systems in which 
the reconstruction of the source occurs in multiple stages. In these systems, a single encoder 
encodes the source and communicates with either one decoder or multiple decoders in 
a step-by-step manner. At each stage, the encoder transmits a portion of the source information 
to the corresponding decoder, which has access also to all previous transmissions. 
Each decoder uses all available transmissions to reconstruct the source, 
possibly incorporating additional side information. The quality of the reconstruction 
at each stage (or by each decoder) is evaluated based on a predefined distortion measure.
One of the motivations of this hierarchical structure is to allow scalability
that meets the available channel resources of the various users who receive
the compressed information or even to adapt to
time-varying channel conditions of a single user. 
Several studies have addressed the successive refinement problem for
probabilistic sources, most notably, memoryless sources, see, e.g.,
\cite{EC91}, where necessary and sufficient conditions for simultaneously 
achieving the rate-distortion function at all stages, \cite{Koshelev94}
and \cite{Rimoldi94}, where the rate-distortion region was fully characterized
in general, and \cite{KN96}, where source coding error exponents were derived. In some later
works, such as \cite{SM04}, \cite{MM08}, \cite{MM10}, and \cite{TD06},
successive refinement coding was considered also with the incorporation
of side information.
Successive-refinement coding is also an important special case of the
so called {\em multiple description coding}, which in its simplest form, consists of
two encoders that send two different individual descriptions of the source to
two separate respective decoders (that do not cooperate with each other), and the compressed bit-streams pertaining to
those descriptions are also
combined and sent to yet
another decoder, whose role is to produce a better reconstruction than both of those of
the two individual decoders. The problem of fully characterizing the
rate-distortion region of the multiple description coding problem is open in
its general form, and there are certain inner and outer bounds to this region, 
see Chapter 13 of \cite{EGK11} for some details and references therein.

In this paper, we focus on successive refinement of information in the context of individual
sequences, namely, deterministic source sequences, as opposed to the
traditional setting of random
sequences governed by a certain probabilistic mechanism. In that sense, this
work can be viewed as an additional step in the development of multiuser information theory for
individual sequences, following a series of earlier works of this flavor,
initiated by Ziv and Lempel in
\cite{LZ76}, \cite{ZL78}, \cite{Ziv80}, and \cite{Ziv84}, and continued by
others in many articles, such as \cite{YK96}, \cite{UK03},
\cite{me00}, \cite{p233}, \cite{p238}, and \cite{mesw}. 
In particular, 
we consider the problem of successive-refinement coding for lossy
compression of individual sequences, in two stages,
where in the first stage, a coarse description at a relatively low rate is
sent from the encoder to the decoder,
and in the second stage, additional coding rate is allocated in order to
refine the description and thereby improve the reproduction. Our main results are in establishing outer bounds
(converse theorems) for the rate region 
with individual sequences, where we
limit the encoders to be finite-state machines similarly as in \cite{ZL78},
\cite{Ziv80}, \cite{Ziv84} and others.
The compatible achievability
scheme is conceptually straightforward, and so, we believe
that the deeper and more interesting part of the contribution of this work is in the
converse theorems, namely, in the outer bounds. Our results are formulated and
proved for two stages of coding, but their extension to any fixed number of
stages is straightforward. These results
can also be viewed as an extension of the fixed-distortion results of
\cite{YK96} to successive-refinement coding.

We also consider the more general multiple
description coding problem and propose achievability schemes that are
analogous to the well-known El Gamal-Cover \cite{EGC82}
and the Zhang-Berger \cite{ZB87} achievability schemes
of memoryless sources and additive distortion measures. There is a clear
parallelism between the rate expressions that we obtain and those of \cite{EGC82}
and \cite{ZB87}, including those that are associated with the gaps between the
outer bound and the corresponding inner bounds.

The outline of the remaining part of this paper is as follows.
In Section \ref{ncpf}, we establish notation conventions and formulate
the problem setting. In Section \ref{background}, we provide some general
background on the LZ algorithm as well as on its conditional form.
Section \ref{sr} is devoted to the above mentioned successive refinement outer
bound, and finally, in Section \ref{mdc}, we 
address the more general multiple description problem.

\section{Notation Conventions and Problem Formulation}
\label{ncpf}

Throughout the paper, random variables will be denoted by capital
letters, specific values they may take will be denoted by the
corresponding lower case letters, and their alphabets
will be denoted by calligraphic letters. Random
vectors, their realizations, and their alphabets will be denoted,
respectively, by capital letters, the corresponding lower case letters, and
the corresponding calligraphic letters, all
superscripted by their dimensions.
More specifically, for a given positive integer, $n$, the source vector $(x_1,x_2,\ldots,x_n)$, with
components, $x_i$, $i=1,2,\ldots,n$, from a
finite-alphabet, $\calX$, will be denoted by $x^n$. The set of all such
$n$-vectors will be denoted by
$\calX^n$, which is the $n$--th order Cartesian power of the single-letter
source alphabet, $\calX$.
Likewise, reproduction vectors of length $n$, such as $(\hx_1,\ldots,\hx_n)$
and $(\tx_1,\ldots,\tx_n)$, with
components, $\hx_i$ and $\tx_i$, $i=1,\ldots,n$, from
finite-alphabets, $\hat{\calX}$ and $\tcalX$, will be denoted by
$\hx^n\in\hat{\calX}^n$ and $\tx^n\in\tcalX^n$, respectively.
An infinite source sequence $(x_1,x_2,\ldots)$ will be denoted by $\bx$.
The cardinalities of $\calX$, $\hcalX$, and $\tcalX$ will be denoted by
$\alpha$, $\beta$, and $\gamma$, respectively. The value $\alpha=\infty$ is
allowed, to incorporate countably infinite and continuous source alphabets.
By contrast, $\beta$ and $\gamma$ will always be finite.

For two positive integers $i$ and $j$, where $i\le j$, the notation $x_i^j$ will be used to denote the substring
$(x_i,x_{i+1},\ldots,x_j)$. For $i=1$, the subscript `1' will be omitted, and
so, the shorthand notation of $(x_1,x_2,\ldots,x_n)$ will be $x^n$ as
mentioned before. Similar
conventions will apply to other sequences.
Probability distributions will be denoted by the letter $P$ with
possible subscripts, depending on the context.
The indicator function for an event $\calA$ will be denoted by
$\calI\{\calA\}$, that is, $\calI\{\calA\}=1$ is $\calA$ occurs, and
$\calI\{\calA\}=0$, if not.
The logarithmic function, $\log x$, will be understood to be defined to the
base 2. Logarithms to the base $e$ will be denote by $\ln$.
Let $d_1:\calX^n\times\hat{\calX}^n\to\reals^+$  and
$d_2:\calX^n\times\tcalX^n\to\reals^+$ 
be two arbitrary distortion functions between
source vectors, $x^n$, and corresponding reproduction vectors, $\hx^n$ and
$\tx^n$, respectively. 

The successive refinement encoder model is as follows. It is composed of a
cascade of two encoders: a reproduction encoder (namely, a vector
quantizer) followed by a lossless encoder.
The input to the reproduction encoder is the source sequence $x^n$ and the
output is a pair of reproduction vectors,
$(\hx^n,\tx^n)\in\hat{\calX}^n\times\tilde{\calX}^n$ that obeys the distortion
constraints, $d_1(x^n,\hx^n)\le nD_1$ and
$d_2(x^n,\tx^n)\le nD_2$, where $D_1\ge 0$ and $D_2\ge 0$ are prescribed
normalized distortion levels. There are no particular restrictions imposed on
the distortion functions. We denote

\begin{equation}
\calB(x^n)=\{(\hx^n,\tx^n):~d_1(x^n,\hx^n)\le nD_1,~d_2(x^n,\tx^n)\le nD_2\}.
\end{equation}
The mapping 
$\calX^n\to\hat{\calX}^n\times\tilde{\calX}^n$, employed by the reproduction
encoder, is arbitrary and not limited. The pair $(\hx^n,\tx^n)$ serves as an
input to the lossless encoder.

As for the lossless encoder,
we follow the same modeling approach as in \cite{ZL78}, but with a few
adjustments to make it suitable to successive refinement. In particular,
the lossless encoder 
is defined by a set

$$E=(\hcalX,\tcalX,\calU,\calV,\calS,\calZ,f_1,f_2,g_1,g_2),$$
where $\hcalX$ and $\tcalX$ are as before;
$\calU$ and $\calV$ are two sets of variable-length
binary strings,
which both include the empty string $\lambda$ 
of length zero; $\calS$ and $\calZ$ are two sets of states, each one
containing $q$ states; $f_1:\calS\times\hcalX\to\calU$ and
$f_2:\calZ\times\hcalX\times\tcalX\to\calV$ are the encoder output functions, and
finally, $g_1:\calS\times\hcalX\to\calS$ and
$g_2:\calZ\times\hcalX\times\tcalX\to\calZ$ are two next-state functions.
When the lossless encoder is fed by a sequence of pairs
$(\hx_1,\tx_1),(\hx_2,\tx_2),\ldots$, it outputs a corresponding sequence of
pairs of binary strings, $(u_1,v_1),(u_2,v_2),\ldots$ according to the following recursive
mechanism. For $t=1,2,\ldots$:

\begin{eqnarray}
u_t&=&f_1(s_t,\hx_t)\\
s_{t+1}&=&g_1(s_t,\hx_t)\\
v_t&=&f_2(z_t,\hx_t,\tx_t)\\
z_{t+1}&=&g_2(z_t,\hx_t,\tx_t),
\end{eqnarray}
where the initial states, $s_1$ and $z_1$, are assumed arbitrary fixed
members of $\calS$ and $\calZ$, respectively. Similarly as in \cite{ZL78}, we adopt
the extended notation, of $f_1(s_1,\hx^n)$ for $u^n$, $g_1(s_1,\hx^n)$ for
$s_{n+1}$, and similar notations associated with $f_2$ and $g_2$.

An encoder $E$ is said to be information lossless if for every positive
integer $k$, the vector $(s_1,f_1(s_1,\hx^k),g_1(s_1,\hx^k))$ uniquely determines $\hx^k$ and
likewise,

$$(s_1,z_1,f_1(s_1,\hx^k),f_2(z_1,\hx^k,\tx^k),g_1(s_1,\tx^k),g_2(z_1,\hx^k,\tx^k))$$ 
uniquely determines $(\hx^k,\tx^k)$.
Let $\calE(q)$ be the set of all information lossless encoders, $\{E\}$, with
$|\calS|\le q$ and $|\calZ|\le q$.

Given a lossless encoder $E$ and a pair of inputs $(\hx^n,\tx^n)$, we define

\begin{equation}
[\rho_E(\hx^n)]_1=\frac{L(u^n)}{n}=\frac{L[f_1(s_1,\hx^n)]}{n}
\end{equation}
and

\begin{equation}
[\rho_E(\hx^n,\tx^n)]_2=\frac{L(u^n)+L(v^n)}{n}=\frac{L[f_1(s_1,\hx^n)]+L[f_2(z_1,\hx^n,\tx^n)]}{n},
\end{equation}
where $L(u^n)=\sum_{i=1}^nl(u_i)$, $l(u_i)$ being the length (in bits) of the
binary string $u_i$, and similarly for $L(v^n)$. Recall that for the empty
string, $\lambda$, we define $l(\lambda)=0$.
The achievable rate region for $x^n$ that is associated with $E$ is defined as

\begin{equation}
\calR_E(x^n)=\bigcup_{(\hx^n,\tx^n)\in\calB(x^n)}\{(R_1,R_2):~R_1\ge
[\rho_E(\hx^n)]_1, R_1+R_2\ge [\rho_E(\hx^n,\tx^n)]_2\},
\end{equation}
and the
$q$-state achievable rate region for $x^n$ is defined as

\begin{equation}
\calR_q(x^n)=\bigcup_{E\in\calE(q)}\calR_E(x^n).
\end{equation}
The rationale behind the union operations in these definitions is that they are
two-dimensional set-theoretic
analogues of the minimization operations over the appropriate encoders $\{E\}$ and
reproduction vectors, $\{\hx^n\}$, that appear for a single coding rate, like in
\cite{ZL78} and \cite{YK96}, as can be noted from the simple relationship:

\begin{eqnarray}
& &\left\{R:~R\ge\min_{E\in\{\mbox{all $q$-state encoders}\}}\min_{\{\hx:~d(x^n,\hx^n)\le
nD\}}\rho_E(\hx^n)\right\}\nonumber\\
&=&\bigcup_{E\in\{\mbox{all $q$-state
encoders}\}}\bigcup_{\{\hx^n:~d(x^n,\hx^n)\le nD\}}\{R:~R\ge \rho_E(\hx^n)\}
\end{eqnarray}
for a given generic distortion function $d$ and distortion level $D$.

For later use, we also define
the joint empirical distribution of $\ell$-blocks of
$(\hx_{i\ell+1}^{i\ell+\ell},\tx_{i\ell+1}^{i\ell+\ell})$, $\ell=0,1,\ldots,n/\ell-1$, provided that $\ell$
divides $n$. Specifically, consider the
empirical distribution,
$\hat{P}=\{\hat{P}(\hx^\ell,\tx^\ell),~\hx^\ell\in\hat{\calX}^\ell,~\tx^\ell\in\tcalX^\ell\}$,
of pairs of $\ell$-vectors, defined as

\begin{equation}
\hat{P}(\hx^\ell,\tx^\ell)=\frac{\ell}{n}\sum_{i=0}^{n/\ell-1}
\calI\{\hx_{i\ell+1}^{i\ell+\ell}=\hx^\ell,~\tx_{i\ell+1}^{i\ell+\ell}=\tx^\ell\},~~~~\hx^\ell\in\hat{\calX}^\ell,~
\tx^\ell\in\tcalX^\ell
\end{equation}
Let $H(\hX^\ell,\tX^\ell)$ denote the joint
empirical entropy of an auxiliary pair of random $\ell$-vectors,
$(\hX^\ell,\tX^\ell)$,
induced by $\hat{P}$, that is,

\begin{equation}
H(\hX^\ell,\tX^\ell)=-\sum_{(\hx^\ell,\tx^\ell)\in\hat{\calX}^\ell\times\tcalX^\ell}
\hat{P}(\hx^\ell,\tx^\ell)\log
\hat{P}(\hx^\ell,\tx^\ell).
\end{equation}
Accordingly, $H(\hX^\ell)$ and $H(\tX^\ell|\hX^\ell)$ will denote the
corresponding marginal empirical entropy of $\hX^\ell$ and the conditional empirical
entropy of $\tX^\ell$ given $\hX^\ell$.

Our objective is to provide inner and outer bounds to the achievable rate
region and show that they asymptotically coincide in the limit of large $n$
followed by a limit of large $q$, in analogy to the asymptotic regime of
\cite{ZL78}. 

\section{Background}
\label{background}

Before the exposition of the main results and their proofs,
we revisit key terms and details
related to the 1978 version of the LZ algorithm, also known as the LZ78
algorithm \cite{ZL78}, which is the central building block in this work.
The incremental parsing procedure of the LZ78 algorithm is a sequential parsing
process applied to a finite-alphabet input vector, $\hx^n$.
According to this procedure, each
new phrase is the shortest string not encountered before as a parsed phrase,
except for the potential incompleteness of the last phrase. For instance, the
incremental parsing of the vector $\hx^{15}=\mbox{abbabaabbaaabaa}$ results in
$\mbox{a,b,ba,baa,bb,aa,ab,aa}$. Let $c(\hx^n)$ denote the
number of phrases in $\hx^n$ resulting from the incremental parsing procedure
(in the above example, $c(\hx^{15})=8$).
Furthermore, let $LZ(\hx^n)$ denote the length of the LZ78 binary compressed
code for $\hx^n$. According to \cite[Theorem 2]{ZL78}, the following inequality
holds:

\begin{eqnarray}
\label{lz-clogc}
LZ(\hx^n)&\le&[c(\hx^n)+1]\log\{2\beta[c(\hx^n)+1]\}\nonumber\\
&=&c(\hx^n)\log[c(\hx^n)+1]+c(\hx^n)\log(2\beta)+\log\{2\beta[c(\hx^n)+1]\}\nonumber\\
&=&c(\hx^n)\log c(\hx^n)+c(\hx^n)\log\left[1+\frac{1}{c(\hx^n)}\right]+
c(\hx^n)\log(2\beta)+\log\{2\beta[c(\hx^k)+1]\}\nonumber\\
&\le&c(\hx^n)\log c(\hx^n)+\log
e+\frac{n(\log \beta)\log(2\beta)}{(1-\varepsilon_n)\log
n}+\log[2\beta(n+1)]\nonumber\\
&\dfn&c(\hx^n)\log c(\hx^n)+n\cdot\epsilon(n),
\end{eqnarray}
where we remind that $\beta$ is the cardinality of $\hcalX$, and where
both $\varepsilon_n$ and $\epsilon(n)$ tends to zero as $n\to\infty$.
In other words, the LZ code-length for $\hx^n$ is upper bounded by
an expression whose main term is $c(\hx^n)\log c(\hx^n)$. On the other hand,
$c(\hx^n)\log c(\hx^n)$ is also known to be the main term of a lower bound
(see Theorem 1 of \cite{ZL78})
to the shortest code-length attainable by any information lossless finite-state encoder with no
more than $q$ states, provided that $\log(q^2)$ is very small compared to
$\log c(\hx^n)$. In view of these facts, we henceforth refer to $c(\hx^n)\log
c(\hx^n)$ as the unnormalized {\em LZ
complexity} of $\hx^n$ whereas the normalized LZ complexity is defined as

\begin{equation}
\rho_{\mbox{\tiny LZ}}(\hx^n)\dfn
\frac{c(\hx^n)\log
c(\hx^n)}{n}.
\end{equation}

A useful inequality, that relates the empirical entropy of non-overlapping
$\ell$-blocks of $\hx^n$ (where $\ell$ divides $n$) and $\rho_{\mbox{\tiny
LZ}}(\hx^n)$ (see, for example,
eq.\ (26) of \cite{me23}), is the following:

\begin{eqnarray}
\label{zivineq}
\frac{H(\hX^\ell)}{\ell}&\ge&\rho_{\mbox{\tiny LZ}}(x^n)
-\frac{\log[4\beta^{2\ell}]\log\beta}{(1-\varepsilon_n)\log
n}-\frac{\beta^{2\ell}\log[4\beta^{2\ell}]}{n}-\frac{1}{\ell}\nonumber\\
&\dfn&\rho_{\mbox{\tiny LZ}}(x^n)-\delta_n(\ell),
\end{eqnarray}
It is obtained from the fact that the Shannon code for $\ell$-blocks can be
implemented using a finite-state encoder with no more than $\beta^\ell$
states. Specifically, for a block code of length $\ell$ to be implemented by a
finite-state machine, one defines the state at each time instant $i$ to be the
contents of the input, starting at the beginning of the current block (at time
$\ell\cdot\lfloor i/\ell\rfloor+1$) and ending at time $i-1$. The number of states
for an input alphabet of size $\beta$ is then
$\sum_{i=0}^{\ell-1}\beta^i=(\beta^\ell-1)/(\beta-1)<\beta^\ell$. Therefore,
the code-length of this Shannon code
must comply with the lower bound of Theorem 1 in \cite{ZL78}.
Note that $\lim_{n\to\infty}\delta_n(\ell)=1/\ell$ and so,
$\lim_{\ell\to\infty}\lim_{n\to\infty}\delta_n(\ell)=0$. Clearly, it is
possible to let $\ell=\ell(n)$ increase with $n$ slowly enough such that
$\delta_n(\ell(n))\to 0$ as $n\to\infty$, in particular, $\ell(n)$ should be $o(\log n)$
for that purpose.

In \cite{Ziv85}, the notion of the LZ complexity was extended to incorporate
finite-state lossless compression in the presence of side information, namely,
the conditional version of the LZ complexity.
Given $\hx^n$ and $\tx^n$,
let us apply the incremental
parsing procedure of the LZ algorithm
to the sequence of pairs $((\hx_1,\tx_1),(\hx_2,\tx_2),\ldots,(\hx_n,\tx_n))$.
As mentioned before, according to this procedure, all phrases are distinct
with a possible exception of the last phrase, which might be incomplete.
Let $c(\hx^n,\tx^n)$ denote the number of distinct phrases.
As an example (taken from \cite{Ziv85}), let $n=6$ and consider the sequence
pair $(\hx^6,\tx^6)$ along with its joint incremental parsing as follows:

\begin{eqnarray}
\hx^6&=&0~|~1~|~0~1~|~0~1|\nonumber\\
\tx^6&=&0~|~1~|~0~0~|~0~1|
\end{eqnarray}
then $c(\hx^6,\tx^6)=4$.
Let $c'(\hx^n)$ denote the resulting number of distinct phrases
of $\hx^k$ (which may differ from $c(\hx^n)$ in individual parsing of $\hx^n$
alone), and let $\hx(l)$ denote the $l$-th distinct $\hx$--phrase,
$l=1,2,\ldots,c'(\hx^n)$. In the above example, $c(\hx^6)=3$. Denote by
$c_l(\tx^n|\hx^n)$ the number of occurrences of $\hx(l)$ in the
parsing of $\hx^n$, or equivalently, the number of distinct $\tx$-phrases
that jointly appear with $\hx(l)$. Clearly, $\sum_{l=1}^{c'(\hx^n)}
c_l(\tx^n|\hx^n)=
c(\hx^n,\tx^n)$. In the above example, $\hx(1)=0$, $\hx(2)=1$, $\hx(3)=01$,
$c_1(\tx^6|\hx^6)=c_2(\tx^6|\hx^6)=1$, and $c_3(\tx^6|\hx^6)=2$. Now, the conditional LZ
complexity of $\tx^n$ given $\hx^n$ is defined as

\begin{equation}
\rho_{LZ}(\tx^n|\hx^n)\dfn\frac{1}{n}\sum_{l=1}^{c'(\hx^n)}c_l(\tx^n|\hx^n)\log
c_l(\tx^n|\hx^n).
\end{equation}
In \cite{Ziv85} it was shown that $\rho_{\mbox{\tiny LZ}}(\tx^n|\hx^n)$ is the main term of the
compression ratio achieved by the conditional version of the LZ algorithm
for compressing $\tx^n$ in the presence of the side information $\hx^n$,
available to both encoder and decoder -- see the compression scheme
described in \cite{Ziv85} (see
also \cite{UK03}), i.e., the length function, $LZ(\tx^n|\hx^n)$, of the coding scheme
proposed therein is upper bounded (in parallel to (\ref{lz-clogc})) by

\begin{equation}
\label{conditional-lz}
LZ(\tx^n|\hx^n)\le n\rho_{LZ}(\tx^n|\hx^n)+n\hat{\epsilon}(n),
\end{equation}
where $\hat{\epsilon}(n)=O\left(\frac{\log(\log n)}{\log n}\right)$ (see eqs.\
(10) and (11) in \cite{UK03}).
On the other hand, analogously to \cite[Theorem 1]{ZL78}, it was shown in
\cite{me00}, that $\rho_{LZ}(\tx^n|\hx^n)$ is also the main term of a lower bound
to the compression ratio that can be achieved by any finite-state encoder with
side information at both ends, provided that the number of states is not
too large, similarly as described above for the unconditional version.

The inequality (\ref{zivineq}) also extends to the conditional case as
follows (see \cite{me00}):

\begin{equation}
\label{conditionalzivineq}
\frac{H(\tX^\ell|\hX^\ell)}{\ell}
\ge\rho_{\mbox{\tiny LZ}}(\tx^n|\hx^n)-\delta_n'(\ell),
\end{equation}
where $\delta_n'(\ell)$ is the same as $\delta_n(\ell)$ except that
$\beta^\ell$ therein is replaced by
$(\beta\gamma)^\ell$ to accommodate the number of
states associated with the conditional version of the aforementioned Shannon code applied to
$\ell$-blocks. By the same token, we also have

\begin{equation}
\frac{H(\hX^\ell,\tX^\ell)}{\ell}
\ge\rho_{\mbox{\tiny LZ}}(\hx^n,\tx^n)-\delta_n'(\ell).
\end{equation}

\section{The Outer Bound for Successive Refinement}
\label{sr}

Our main result for finite-state encoders is the following.\\

\noindent 
{\em Theorem 1.} 
For every $x^n\in\calX^n$,

\begin{eqnarray}
\calR_q(x^n)&\subseteq&\calR_{\mbox{\tiny o}}(x^n)\nonumber\\
&\dfn&\bigcup_{(\hx^n,\tx^n)\in\calB(x^n)}\calR_{\mbox{\tiny LZ}}(\hx^n,\tx^n),
\end{eqnarray}
where

\begin{eqnarray}
\calR_{\mbox{\tiny LZ}}(\hx^n,\tx^n)&\dfn&\bigg\{(R_1,R_2):\nonumber\\
R_1&\ge&\rho_{\mbox{\tiny LZ}}(\hx^n) - \Delta_1(q,n),\nonumber\\
R_1+R_2&\ge&\rho_{\mbox{\tiny LZ}}(\hx^n)+\rho_{\mbox{\tiny LZ}}(\tx^n|\hx^n)
-\Delta_2(q,n)\bigg\},
\end{eqnarray}
where $\Delta_1(q,n)$ and $\Delta_2(q,n)$ are defined as

\begin{equation}
\Delta_1(q,n)=\frac{\log(4q^2)\log\beta}{(1-\epsilon_n)\log n}+\frac{q^2\log(4q^2)}{n}
\end{equation}
with $\epsilon_n\to 0$ as $n\to\infty$, and

\begin{eqnarray}
\Delta_2(n,q)&=&\min_{\{\ell:~\ell~\mbox{\tiny divides}~n\}}
\bigg\{\delta_n(\ell)+\delta_n'(\ell)+
\frac{1}{\ell}\log\left[q^4\left(1+\log\left[1+\frac{\beta^\ell\gamma^\ell}{q^4}\right]\right)\right]\bigg\}.
\end{eqnarray}

\noindent
{\bf Discussion.} Several comments are now in order.\\

\noindent
1. Since both $\Delta_1(q,n)$ and $\Delta_2(q,n)$ tend to zero as $n\to\infty$ for fixed
$q$, the asymptotic achievability of $\calR_{\mbox{\tiny o}}(x^n)$ 
is conceptually straightforward: Given an internal point, $(R_1,R_2)\in
\calR_{\mbox{\tiny o}}(x^n)$, there must be at least one pair
$(\hx^n,\tx^n)\in\calB(x^n)$ such that
$(R_1,R_2)\in\calR_{\mbox{\tiny LZ}}(\hx^n,\tx^n)$. Upon finding such a
pair, proceed as follows: At the first stage, apply LZ78 compression to $\hx^n$ at a coding rate of
$R_1=\frac{LZ(\hx^n)}{n}$ which is only slightly above $\rho_{\mbox{\tiny
LZ}}(\hx^n)$ for large $n$, as discussed in Section \ref{background}.
At the second stage, apply conditional LZ compression of $\tx^n$ given $\hx^n$
as side information at both ends, at an incremental coding rate of 
$R_2=\frac{LZ(\tx^n|\hx^n)}{n}$ which is close to $\rho_{\mbox{\tiny
LZ}}(\tx^n|\hx^n)$, as also explained in Section \ref{background}, and then
the total rate, $R_1+R_2$, is about $\rho_{\mbox{\tiny LZ}}(\hx^n)+\rho_{\mbox{\tiny
LZ}}(\tx^n|\hx^n)$. Similarly as in \cite{ZL78}, there is still a certain gap 
between the achievability and the converse theorem, because the
achievability requires encoders whose number of states is not small compared
to $n$, whereas the converse is significant when $q$ is very small relative to
$n$. As in \cite{ZL78}, this gap can be closed in the asymptotic limit of
large $q$ by partitioning
the sequence into non-overlapping blocks and starting
over the LZ compression mechanism in each block separately. We will address
this point in detail later on.\\

\noindent
2. Considering that the achievability is conceptually straightforward, as explained in
item no.\ 1 above, the interesting and deeper result is the converse
theorem. Since the second stage encoder receives both $\hx^n$ and $\tx^n$ as
inputs, it is immediate to lower bound
the total coding rate, at the second stage, in terms of the joint compressibility of
$(\hx^n,\tx^n)$, namely by $\rho_{\mbox{\tiny LZ}}(\hx^n,\tx^n)$, but recall
that the first-stage encoder must have already allocated a rate at least as large as
$\rho_{\mbox{\tiny LZ}}(\hx^n)$, then in order to meet a lower bound of $\rho_{\mbox{\tiny LZ}}(\hx^n,\tx^n)$
on the total coding rate, the incremental rate, $R_2$, of the second stage must
not exceed $\rho_{\mbox{\tiny LZ}}(\hx^n,\tx^n)-\rho_{\mbox{\tiny
LZ}}(\hx^n)$, and there is no apparent way to achieve such a coding rate, as
far as the author can see. Nonetheless, since we can also lower
bound the total rate of both stages by $\rho_{\mbox{\tiny
LZ}}(\hx^n)+\rho_{\mbox{\tiny LZ}}(\tx^n|\hx^n)$, then the achievability
becomes obvious, as said. This point is not trivial
because there is no chain rule that applies to the LZ
complexities of arbitrary finite sequences. 
The proof that $\rho_{\mbox{\tiny LZ}}(\hx^n)+\rho_{\mbox{\tiny
LZ}}(\tx^n|\hx^n)$ also serves as a lower bound (essentially), requires a
certain manipulation by using a generalized Kraft inequality and passing via
empirical entropies, as can be seen in the proof.\\

\noindent
3. The choice of $\hx^n$ exhibits a trade-off between the coding rate of the
first stage and the incremental rate at the second stage because $\hx^n$ is
both compressed at the first stage and serves as side information at the second stage,
so there might be a certain tension between selecting $\hx^n$ for having
small $\rho_{\mbox{\tiny LZ}}(\hx^n)$ and selecting it for small
$\rho_{\mbox{\tiny LZ}}(\tx^n|\hx^n)$. Of course, an analogous tension exists also in
successive refinement for memoryless sources \cite{Rimoldi94}. The
reproduction encoder must select $(\hx^n,\tx^n)\in\calB(x^n)$ that best
compromises these criteria.\\

\noindent
4. The results extend straightforwardly to any finite number of stages, where
at each stage one applies conditional LZ compression of the current
reproduction given all previous reproductions.\\

\noindent
{\em Proof of Theorem 1.} 
We begin from the first stage.
By definition, if $(R_1,R_2)\in\calR_q(x^n)$, then
there must exist an encoder $E\in\calE(q)$ and
$(\hx^n,\tx^n)\in\calB(x^n)$ 
such that $R_1\ge[\rho_E(\hx^n)]_1$
and $R_1+R_2\ge [\rho_E(\hx^n,\tx^n)]_2$. Now, according to Theorem 1 of
\cite{ZL78}:

\begin{eqnarray}
[\rho_E(\hx^n)]_1&\ge&
\frac{c(\hx^n)+q^2}{n}\cdot\log\left[\frac{c(\hx^n)+q^2}{4q^2}\right]+\frac{2q^2}{n}\nonumber\\
&>&\frac{c(\hx^n)+q^2}{n}\cdot\log[c(\hx^n)+q^2]-\frac{c(\hx^n)+q^2}{n}\log(4q^2)\nonumber\\
&>&\frac{c(\hx^n)\log c(\hx^n)}{n}-\frac{c(\hx^n)\log(4q^2)}{n}-\frac{q^2\log(4q^2)}{n}\nonumber\\
&\ge&\frac{c(\hx^n)\log c(\hx^n)}{n}-\frac{\log(4q^2)\log\beta}{(1-\epsilon_n)\log n}-\frac{q^2\log(4q^2)}{n}\nonumber\\
&=&\rho_{\mbox{\tiny LZ}}(\hx^n)-\Delta_1(q,n),
\end{eqnarray}
where $\lim_{n\to\infty}\epsilon_n=0$ and the last inequality is an
application of eq.\ (6) in \cite{ZL78}. Since $R_1\ge[\rho_E(\hx^n)]_1$, it follows that
\begin{equation}
R_1\ge \rho_{\mbox{\tiny LZ}}(\hx^n)-\Delta_1(q,n).
\end{equation}

Moving on to the combined encoder of both stages, consider the following.
According to Lemma 2 of \cite{ZL78} and due to the
postulated information losslessness, the combined encoder, which has
$q^2$ states, must obey the following generalized Kraft
inequality:

\begin{equation}
\sum_{(\hat{x}^\ell,\tx^\ell\in\hat{\calX}^\ell\times\tcalX^\ell}
\exp_2\left\{-[\min_{s\in\calS}L[f_1(s,\hat{x}^\ell)]+\min_{z\in\calZ}L[f_2(z,\hx^\ell,\tx^\ell)]\right\}\le
q^4\left(1+\log\left[1+\frac{\beta^\ell\gamma^\ell}{q^4}\right]\right).
\end{equation}
This implies
that the description length at the output of this encoder is
lower bounded as follows.

\begin{eqnarray}
n(R_1+R_2)&\ge&n[\rho_E(\hx^n,\tx^n)]_2\nonumber\\
&=&L(u^n)+L(v^n)\nonumber\\
&=&\sum_{t=1}^n \{L[f_1(s_t,\hat{x}_t)]+L[f_2(z_t.\hx_t,\tx_t)]\}\nonumber\\
&=&\sum_{m=0}^{n/\ell-1} \sum_{j=1}^\ell
\{L[f_1(s_{m\ell+j},\hat{x}_{m\ell+j})]+
L[f_2(z_{m\ell+j},\hat{x}_{m\ell+j}),\tx_{m\ell+j})]\}\nonumber\\
&=&\sum_{m=0}^{n/\ell-1}
\{L[f_1(s_{m\ell+1},\hat{x}_{m\ell+1}^{m\ell+\ell})]+
L[f_2(z_{m\ell+1},\hat{x}_{m\ell+1}^{m\ell+\ell},\tx_{m\ell+1}^{m\ell+\ell})]\}\nonumber\\
&\ge&\sum_{m=0}^{n/\ell-1}
\left\{\min_{s\in\calS}L[f_1(s,\hat{x}_{m\ell+1}^{m\ell+\ell})]+
\min_{z\in\calZ}L[f_2(z,\hat{x}_{m\ell+1}^{m\ell+\ell},\tx_{m\ell+1}^{m\ell+\ell})]\right\}\nonumber\\
&=&\frac{n}{\ell}\sum_{(\hat{x}^\ell,\tx^\ell)\in\hat{\calX}^\ell\times\tcalX^\ell}\hat{P}(\hat{x}^\ell,\tx^\ell)
\cdot\left\{\min_{s\in\calS}L[f_1(s,\hat{x}^\ell)]+\min_{z\in\calZ}L[f_2(z,\hat{x}^\ell,\tx^\ell)]\right\}
\end{eqnarray}
and so,

\begin{equation}
R_1+R_2\ge\frac{1}{\ell}\sum_{(\hat{x}^\ell,\tx^\ell\in\hat{\calX}^\ell\times\tcalX^\ell}\hat{P}(\hat{x}^\ell,\tx^\ell)
\cdot\left\{\min_{s\in\calS}L[f_1(s,\hat{x}^\ell)]+
\min_{z\in\calZ}L[f_2(z,\hat{x}^\ell,\tx^\ell)]\right\}.
\end{equation}
Now, by the generalized Kraft inequality above,

\begin{eqnarray}
& &q^4\left(1+\log\left[1+\frac{\beta^\ell\gamma^\ell}{q^4}\right]\right)\nonumber\\
&\ge&\sum_{(\hat{x}^\ell,\tx^\ell)\in\hat{\calX}^\ell\times\tcalX^\ell}
\exp_2\left\{-\left(\min_{s\in\calS}L[f_1(s,\hat{x}^\ell)]+
\min_{z\in\calZ}L[f_2(z,\hat{x}^\ell,\tx^\ell)]\right)
\right\}\nonumber\\
&=&\sum_{(\hat{x}^\ell,\tx^\ell)\in\hat{\calX}^\ell\times\tcalX^\ell}
\hat{P}(\hat{x}^\ell,\tx^\ell)\cdot
\exp_2\left\{-\left(\min_{s\in\calS}L[f_1(s,\hat{x}^\ell)+\min_{z\in\calZ}L[f_2(z,\hat{x}^\ell,\tx^\ell)]\right)-\log
\hat{P}(\hat{x}^\ell,\tx^\ell)\right\}\nonumber\\
&\ge&\exp_2\left\{-\sum_{(\hat{x}^\ell,\tx^\ell)\in\hat{\calX}^\ell\times\tcalX^\ell}\hat{P}(\hat{x}^\ell,\tx^\ell)\cdot
\left(\min_{s\in\calS}L[f_1(s,\hat{x}^\ell)+
[\min_{z\in\calZ}L[f_2(z,\hat{x}^\ell,\tx^\ell)]\right)+H(\hat{X}^\ell,\tX^\ell)\right\},\nonumber
\end{eqnarray}
where the last inequality follows from the convexity of the exponential
function and Jensen's inequality.
This yields

\begin{eqnarray}
& &\log\left\{q^4\left(1+\log\left[1+\frac{\beta^\ell\gamma^\ell}{q^4}\right]\right)\right\}\nonumber\\
&\ge&H(\hat{X}^\ell,\tX^\ell)-
\sum_{(\hat{x}^\ell,\tx^\ell)\in\hat{\calX}^\ell\times\tcalX^\ell}\hat{P}(\hat{x}^\ell,\hx^\ell)\cdot
\left\{\min_{s\in\calS}L[f_1(s,\hat{x}^\ell)]+
\min_{z\in\calZ}L[f_2(z,\hat{x}^\ell,\tx^\ell)]\right\},
\end{eqnarray}
implying that

\begin{eqnarray}
R_1+R_2&\ge&\frac{L(u^n)+L(v^n)}{n}\nonumber\\
&\ge&\frac{1}{\ell}\sum_{(\hat{x}^\ell,\tx^\ell)\in\hat{\calX}^\ell\times\tcalX^\ell}
\hat{P}(\hat{x}^\ell,\tx^\ell)\cdot\left\{\min_{s\in\calS}L[f_1(s,\hat{x}^\ell)]+\min_{z\in\calZ}L[f_2(z,\hat{x}^\ell,\tx^\ell)]\right\}\nonumber\\
&\ge&\frac{H(\hat{X}^\ell,\tX^\ell)}{\ell}-
\frac{1}{\ell}\log\left\{q^4\left(1+\log\left[1+\frac{\beta^\ell\gamma^\ell}{q^4}
\right]\right)\right\}\nonumber\\
&=&\frac{H(\hat{X}^\ell)}{\ell}+
\frac{H(\tX^\ell|\hat{X}^\ell)}{\ell}-
\frac{1}{\ell}\log\left\{q^4\left(1+\log\left[1+\frac{\beta^\ell\gamma^\ell}{q^4}
\right]\right)\right\}\nonumber\\
\end{eqnarray}
Now, according to eq.\ (\ref{zivineq}),

\begin{equation}
\frac{H(\hat{X}^\ell)}{\ell}\ge \rho_{\mbox{\tiny LZ}}(\hx^n)
-\delta_n(\ell).
\end{equation}
Similarly, according to eq.\ (\ref{conditionalzivineq}),

\begin{equation}
\frac{H(\tilde{X}^\ell|\hat{X}^\ell)}{\ell}\ge
\rho_{\mbox{\tiny LZ}}(\tx^n|\hx^n)-\delta_n'(\ell),
\end{equation}
and so,

\begin{eqnarray}
R_1+R_2&\ge&\rho_{\mbox{\tiny LZ}}(\hx^n)+\rho_{\mbox{\tiny
LZ}}(\tx^n|\hx^n)-\delta_n(\ell)-\delta_n'(\ell)-
\frac{1}{\ell}\log\left[q^4\left(1+\log\left[1+\frac{\beta^\ell\gamma^\ell}{q^4}\right]\right)\right].
\end{eqnarray}
Maximizing this lower bound w.r.t.\ $\ell$ yields

\begin{equation}
R_1+R_2\ge \rho_{\mbox{\tiny LZ}}(\hx^n)+\rho_{\mbox{\tiny LZ}}(\tx^n|\hx^n)
-\Delta_2(n,q),
\end{equation}
where

\begin{eqnarray}
\Delta_2(n,q)&=&\min_{\{\ell:~\ell~\mbox{\tiny divides}~n\}}
\bigg\{\delta_n(\ell)+\delta_n'(\ell)+
\frac{1}{\ell}\log\left[q^4\left(1+\log\left[1+\frac{\beta^\ell\gamma^\ell}{q^4}\right]\right)\right]\bigg\}.
\end{eqnarray}
This completes the proof of Theorem 1.\\

Referring to the last part of comment no.\ 1 in the discussion that follows
Theorem 1, we
now address the gap in terms of the number of states.
For an infinite source sequence $\bx=(x_1,x_2,\ldots)$,
we define the $q$-state achievable rate region for $\bx$ as

\begin{equation}
\calR_q(\bx)=\bigcup_{m\ge 1}\bigcap_{n\ge m}\calR_q(x^n),
\end{equation}
and finally, the finite-state achievable rate region for $\bx$ is defined as

\begin{equation}
\calR_\infty(\bx)=\bigcup_{q\ge 1}\calR_q(\bx).
\end{equation}
These definitions are two-dimensional counterparts of eqs.\ (2)--(4) in
\cite{ZL78},
where the finite-state (lossless) compressibility of $\bx$ is defined in several steps.
In particular, the union over intersections in the definition of
$\calR_q(\bx)$ is the set-theoretic analogue of the limit superior operation,
and the union operation in the definition of $\calR_\infty(\bx)$ is parallel
to a limit of $q\to\infty$.

Let $k$ be a positive integer that divides $n$ and consider the partition of
$\hx^n$ and $\tx^n$ into $n/k$ blocks of length $k$, i.e.,
$\hx_{kt+1}^{kt+k}=(\hx_{kt+1},\hx_{kt+2},\ldots,\hx_{kt+k})$ and
$\tx_{kt+1}^{kt+k}=(\tx_{kt+1},\tx_{kt+2},\ldots,\tx_{kt+k})$,
$t=0,1,\ldots,n/k-1$.
Next, define:

\begin{eqnarray}
\calR_{-}^k(\hx^n,\tx^n)&=&\bigg\{(R_1,R_2):~
R_1\ge\frac{k}{n}\sum_{t=0}^{n/k-1}\rho_{\mbox{\tiny LZ}}(\hx_{kt+1}^{kt+k})
-\Delta_1(q,k),\nonumber\\
& &R_1+R_2\ge\frac{k}{n}\sum_{t=0}^{n/k-1}[\rho_{\mbox{\tiny
LZ}}(\hx_{kt+1}^{kt+k})+
\rho_{\mbox{\tiny LZ}}(\tx_{kt+1}^{kt+k}|\hx_{kt+1}^{kt+k})]
-\Delta_2(q,k)\bigg\}.
\end{eqnarray}
Then, similarly as in Theorem 1,

\begin{equation}
\calR_q(x^n)\subseteq
\calR_{\mbox{\tiny o}}^k(x^n)\dfn\bigcup_{(\hx^n,\tx^n)\in\calB(x^n)}
\calR_{-}^k(\hx^n,\tx^n),
\end{equation}
and so, for every positive integer $N$:

\begin{equation}
\bigcap_{n\ge N}\calR_q(x^n)\subseteq\bigcap_{n\ge N}
\calR_{\mbox{\tiny o}}^k(x^n),
\end{equation}
implying that

\begin{equation}
\calR_q(\bx)=\bigcup_{N\ge 1}\bigcap_{n\ge N}\calR_q(x^n)\subseteq
\bigcup_{N\ge 1}\bigcap_{n\ge N}\calR_{\mbox{\tiny
o}}^k(x^n)\dfn\calR_{\mbox{\tiny o}}^k(\bx).
\end{equation}
Since this holds for every positive integer $k$, then

\begin{equation}
\calR_q(\bx)\subseteq\bigcup_{K\ge 1}\bigcap_{k\ge K}\calR_{\mbox{\tiny
o}}^k(\bx)\dfn\calR_{\mbox{\tiny o}}(\bx),
\end{equation}
and so,

\begin{equation}
\calR_\infty(\bx)\subseteq\calR_{\mbox{\tiny o}}(\bx),
\end{equation}
which establishes an asymptotic version of the converse theorem.

As for the direct part, considering the fact that
a block code of length $k$, operating on $k$-tuples of the two reconstruction
vectors can be implemented by a finite-state machine
with no more than $(\beta\gamma)^k$ states, we have

\begin{eqnarray}
\calR_{(\beta\gamma)^k}(x^n)&\supseteq&\calR_{\mbox{\tiny
i}}^k(x^n)\nonumber\\
&\dfn&\bigcup_{(\hx^n,\tx^n)\in\calB(x^n)}\calR_{
+}^k(\hx^n,\tx^n),
\end{eqnarray}
where

\begin{eqnarray}
\calR_{
+}^k(\hx^n,\tx^n)&=&\bigg\{(R_1,R_2):~R_1\ge\frac{k}{n}\sum_{t=0}^{n/k-1}\rho_{\mbox{\tiny
LZ}}(\hx_{kt+1}^{kt+k})
+O\left(\frac{1}{\log k}\right),\nonumber\\
& &R_1+R_2\ge \frac{k}{n}\sum_{t=0}^{n/k-1}\bigg[\rho_{\mbox{\tiny
LZ}}(\hx_{kt+1}^{kt+k})+\nonumber\\
& &\rho_{\mbox{\tiny LZ}}(\tx_{kt+1}^{kt+k}|\hx_{kt+1}^{kt+k}))
\bigg]+O\left(\frac{\log(\log k)}{\log
k}\right)\bigg\}.
\end{eqnarray}

\begin{eqnarray}
\calR_{(\beta\gamma)^k}(\bx)&=&\bigcup_{N\ge 1}\bigcap_{n\ge N}
\calR_{(\beta\gamma)^k}(x^n)\nonumber\\
&\supseteq& \bigcup_{N\ge 1}\bigcap_{n\ge N}\calR_{\mbox{\tiny
i}}^k(x^n)\nonumber\\
&=&\calR_{\mbox{\tiny i}}^k(\bx)\nonumber\\
&\supseteq&\bigcap_{K\ge k}\calR_{\mbox{\tiny i}}^{K}(\bx)\nonumber\\
\end{eqnarray}
and so,

\begin{equation}
\calR_\infty(\bx)=\bigcup_{k\ge 1}\calR_{(\beta\gamma)^k}(\bx)\supseteq
\bigcup_{k\ge 1}\bigcap_{K\ge k}\calR_{\mbox{\tiny
i}}^{K}(\bx)=\calR_{\mbox{\tiny i}}(\bx).
\end{equation}

We have just proved the following theorem:\\

\noindent
{\bf Theorem 2.} For every infinite individual sequence $\bx=(x_1,x_2,\ldots)$,

\begin{equation}
\calR_{\mbox{\tiny o}}(\bx)\supseteq\calR_\infty(\bx)\supseteq
\calR_{\mbox{\tiny i}}(\bx).
\end{equation}

These inner and outer bounds are tight in the sense that the definitions of
$\calR_{\mbox{\tiny o}}(\bx)$ and
$\calR_{\mbox{\tiny i}}(\bx)$ are based on the same building blocks and the
only difference is in terms that tend to zero as $k\to\infty$.

\section{Multiple Description Coding}
\label{mdc}

Consider next the configuration 
that is associated with the multiple description problem (see, e.g., Chap.\ 13
of \cite{EGK11}), where the source is an individual sequence and the encoders
are modeled as finite-state machines. In particular, there are two
$q$-state encoders and three decoders defined as follows.
Encoders 1 and 2 are fed by $x^n$ and produce two reconstructions,
$\hx^n$ and $\tx^n$, with distortions $d_1(x^n,\hx^n)\le nD_1$ and
$d_2(x^n,\tx^n)\le nD_2$, respectively. Encoder 1 then
compresses $\hx^n$ losslessly
and sends a compressed description to Decoder 1. Likewise, Encoder 2 does
the same with $\tx^n$ and sends a compressed form to Decoder 2.
There is no collaboration between Decoders 1 and 2.
The third decoder, Decoder 0, receives both compressed descriptions
and generates yet another reconstruction, $\gx^n$, with distortion
$d_0(x^n,\gx^n)\le nD_0$. 

Using the same technique as in the proof of Theorem 1,
it is easy to prove the following outer bound to the achievable rate
region:

\begin{equation}
\calR_{\mbox{\tiny o}}(x^n)=
\bigcup_{(\hx^n,\tx^n,\gx^n)\in\calB(x^n)}\calR_{\mbox{\tiny
LZ}}(\hx^n,\tx^n,\gx^n),
\end{equation}
where $\calB(x^n)$ is redefined as

\begin{equation}
\calB(x^n)=\left\{(\hx^n,\tx^n,\gx^n):~d_0(x^n\gx^n)\le nD_0,~
d_1(x^n\hx^n)\le nD_1,~
d_2(x^n\tx^n)\le nD_2\right\},
\end{equation}
and

\begin{eqnarray}
\calR_{\mbox{\tiny
LZ}}(\hx^n,\tx^n,\gx^n)&=&\bigg\{(R_1,R_2):~R_1\ge\rho_{\mbox{\tiny
LZ}}(\hx^n)
-\Delta_1(q,n),~R_2\ge\rho_{\mbox{\tiny LZ}}(\tx^n)
-\Delta_1(q,n),\nonumber\\
& &R_1+R_2\ge
\rho_{\mbox{\tiny LZ}}(\gx^n|\hx^n,\tx^n)
+\nonumber\\
& &\rho_{\mbox{\tiny LZ}}(\hx^n,\tx^n)
-\Delta_2(q,n)\bigg\},
\end{eqnarray}
with $\Delta_1(q,n)$ and $\Delta_2(q,n)$ being defined
similarly as before. The sum-rate inequality is obtained by considering that the two
encoders together compress losslessly the triple $(\hx^n,\tx^n,\gx^n)$ and so,
the main term of the lower bound to $R_1+R_2$ is the joint empirical entropy of
of $(\hx^n,\tx^n,\gx^n)$, which can be decomposed as the sum of the joint empirical
entropy of $(\hx^n,\tx^n)$ and the conditional empirical entropy of $\gx^n$
given $(\hx^n,\tx^n)$, which in turn are essentially further lower bounded by 
$\rho_{\mbox{\tiny LZ}}(\hx^n,\tx^n)$ and $\rho_{\mbox{\tiny
LZ}}(\gx^n|\hx^n,\tx^n)$,
respectively.

We next present two inner bounds, where the first one is analogous to
the El Gamal-Cover inner bound \cite{EGC82} and the second follows the same
line of thought as that of the Zhang-Berger inner bound \cite{ZB87}.

The former inner bound is given by

\begin{equation}
\calR_{\mbox{\tiny i}}^{\mbox{\tiny EGC}}(x^n)=
\bigcup_{(\hx^n,\tx^n,\gx^n)\in\calB(x^n)}\calR_{\mbox{\tiny
i}}(\hx^n,\tx^n,\gx^n),
\end{equation}
where

\begin{eqnarray}
\calR_{\mbox{\tiny
i}}(\hx^n,\tx^n,\gx^n)&=&\bigg\{(R_1,R_2):~R_1\ge\rho_{\mbox{\tiny LZ}}(\hx^n)
+\epsilon(n),~R_2\ge \rho_{\mbox{\tiny LZ}}(\tx^n)
+\epsilon(n),\nonumber\\
& &R_1+R_2\ge\rho_{\mbox{\tiny LZ}}(\gx^n|\hx^n,\tx^n)+\rho_{\mbox{\tiny
LZ}}(\hx^n,\tx^n)+
\hat{I}(\hx^n;\tx^n)+\epsilon(n)+\hat{\epsilon}(n)\bigg\},
\end{eqnarray}
where $\epsilon(n)$ and $\hat{\epsilon(n)}$ are as in
(\ref{lz-clogc}) and (\ref{conditional-lz}), respectively.

\begin{equation}
\hat{I}(\hx^n;\tx^n)=\rho_{\mbox{\tiny LZ}}(\hx^n)+\rho_{\mbox{\tiny
LZ}}(\tx^n)-\rho_{\mbox{\tiny LZ}}(\hx^n,\tx^n).
\end{equation}
The quantity $\hat{I}(\hx^n;\tx^n)$ plays a role of an empirical 
mutual information between $\hx^n$ and $\tx^n$, which manifests the gap
between the lower bounds to the sum-rate inequalities of the inner bound and
the outer bound, analogously to the mutual information term of the El
Gamal-Cover achievable region. 

The achievability of the above inner bound is as follows.
Given an internal point in $\calR_{\mbox{\tiny i}}^{\mbox{\tiny EGC}}(x^n)$, there must exist 
a reconstruction triple $(\hx^n,\tx^n,\gx^n)$ that meets the distortion
constraints and the corresponding rate inequalities. The encoder applies
individual LZ compression for both $\hx^n$ and $\tx^n$ and sends the
compressed versions, at rates $\rho_{\mbox{\tiny LZ}}(\hx^n)$ and
$\rho_{\mbox{\tiny LZ}}(\tx^n)$ (up to negligibly small terms for large
$n$), to Decoder 1
and Decoder 2, respectively. It then applies conditional LZ compression of
$\gx^n$ given $(\hx^n,\tx^n)$ at rate $\rho_{\mbox{\tiny
LZ}}(\gx^n|\hx^n,\tx^n)$
(up to small terms), and splits this compressed bit
stream between Decoders 1 and 2 without violating their rate
inequalities. The rate sum is then essentially

\begin{eqnarray}
& &\rho_{\mbox{\tiny LZ}}(\gx^n|\hx^n,\tx^n)
+\rho_{\mbox{\tiny LZ}}(\hx^n)
+\rho_{\mbox{\tiny LZ}}(\tx^n)\nonumber\\
&=&\rho_{\mbox{\tiny LZ}}(\gx^n|\hx^n,\tx^n)+\rho_{\mbox{\tiny LZ}}(\hx^n,\tx^n)
+\hat{I}(\hx^n;\tx^n).
\end{eqnarray}

In the above description, we explained that the bit-stream associated with the
conditional compression of $\gx^n$ given $(\hx^n,\tx^n)$ is split between
Decoders 1 and 2 without violating the rate inequalities. This is always
possible because of the following simple fact:
Given an internal point in the region $\calR=\{(R_1,R_2):~R_1>
A,~R_2>B,~R_1+R_2\ge A+B+C\}$, there must exist $0\le D\le C$
such that $(A+D,B+C-D)\in\calR$. In particular, let $D=R_1-A\ge 0$.
Then, $R_2=B+C-D\ge B$ and $R_1+R_2=(A+D)+(B+C-D)=A+B+C$.
In our case, $A=\rho_{\mbox{\tiny LZ}}(\hx^n)$,
$B=\rho_{\mbox{\tiny LZ}}(\tx^n)$, and 
$C=\rho_{\mbox{\tiny LZ}}(\gx^n|\hx^n,\tx^n)$. 

The second achievability scheme, in the spirit of the Zhang-Berger scheme, is as follows.
Here, in addition to $\hx^n$, $\tx^n$ and $\gx^n$, we
also generate an auxiliary finite-alphabet sequence, $u^n$. The encoder applies
LZ compression to $u^n$ and
conditional LZ compression of $\hx^n$ given $u^n$ and sends both bit-streams to Decoder 1.
At the same time, it also applies conditional LZ compression of $\tx^n$ given
$u^n$ and sends the compressed forms of $u^n$ and $\tx^n$ to Decoder 2.
Finally, the encoder applies conditional LZ compression of $\gx^n$ given
$(\hx^n,\tx^n,u^n)$ and splits the compressed bit-stream between Decoders and
Decoder 2 in a manner that meets the rate constraints.
Thus,

\begin{eqnarray}
R_1&\approx&\rho_{\mbox{\tiny LZ}}(u^n)+\rho_{\mbox{\tiny LZ}}(\hx^n|u^n)+
\alpha\rho_{\mbox{\tiny LZ}}(\gx^n|\hx^n,\tx^n,u^n)\\
R_2&\approx&\rho_{\mbox{\tiny LZ}}(u^n)+\rho_{\mbox{\tiny LZ}}(\tx^n|u^n)+
(1-\alpha)\rho_{\mbox{\tiny LZ}}(\gx^n|\hx^n,\tx^n,u^n),
\end{eqnarray}
where $\alpha\in[0,1]$.
Thus,

\begin{eqnarray}
R_1+R_2&\approx&2\cdot
\rho_{\mbox{\tiny LZ}}(u^n)+\rho_{\mbox{\tiny LZ}}(\hx^n|u^n)+
\rho_{\mbox{\tiny LZ}}(\tx^n|u^n)+\rho_{\mbox{\tiny LZ}}(\gx^n|\hx^n,\tx^n,u^n)\nonumber\\
&=&2\cdot
\rho_{\mbox{\tiny LZ}}(u^n)+\rho_{\mbox{\tiny LZ}}(\hx^n,\tx^n|u^n)+
\rho_{\mbox{\tiny LZ}}(\gx^n|\hx^n,\tx^n,u^n)+\hat{I}(\hx^n;\tx^n|u^n),
\end{eqnarray}
where

\begin{equation}
\hat{I}(\hx^n;\tx^n|u^n)=\rho_{\mbox{\tiny LZ}}(\hx^n|u^n)+
\rho_{\mbox{\tiny LZ}}(\tx^n|u^n)-\rho_{\mbox{\tiny LZ}}(\hx^n,\tx^n|u^n),
\end{equation}
is analogous to conditional mutual information.
The first term in the rate sum is analogous to $2I(U;X)$, the sum of the
second and the third is analogous to $I(X;\hX,\tX,\gX|U)$ and the last term is
analogous to $I(\hX;\tX|U)$ (see Theorem 13.4, page
332 in \cite{EGK11}).


\begin{thebibliography}{10}

\bibitem{EC91}
Equitz,~W.~H.~R.; Cover~T.~M.~``Successive refinement of information,'' 
{\em IEEE Trans.~Inform.~Theory\/}, {\bf 1991},
vol.~IT--37, no.~2, pp.~269--275.

\bibitem{Koshelev94}
Koshelev,~V.~N.~``On the divisibility of discrete sources with an 
additive single-letter distortion measure,''
{\em Problems of Information Transmission (IPPI)\/}, {\bf 1994},
vol.~30, no.~1, pp.~27--43.

\bibitem{Rimoldi94}
Rimoldi,~B,~``Successive refinement of information: characterization of 
achievable rates,''
{\em IEEE Trans.~Inform.~Theory\/}, {\bf 1994},
vol.~IT--40, no.~1, pp.~253--259.

\bibitem{KN96}
Kanlis,~A.; Narayan,~P.~``Error exponents for successive refinement 
by partitioning,''
{\em IEEE Trans.~Inform.~Theory\/}, {\bf 1996},
vol.~IT--42, no.~1.

\bibitem{SM04}
Steinberg, Y.; Merhav,~N.~ ``On successive refinement for
the Wyner--Ziv problem,''
{\it IEEE Trans.\ Inform.\ Theory}, {\bf 2004},
vol.\ 50, no.\ 8, pp.\ 1636--1654.

\bibitem{MM08}
Maor, A.;~Merhav, N.~``On successive refinement
with causal side information at the decoders,''
{\it IEEE Trans.\ Inform.\ Theory}, {\bf 2008},
vol.\ 54, no.\ 1, pp.\ 332--343.

\bibitem{MM10}
Maor (Varshavsky),~A; Merhav,~N.~
``On successive refinement for the Kaspi/Heegard--Berger problem,''
{\it IEEE Trans.\ Inform.\ Theory}, {\bf 2010}, vol.\ 56, no.\ 8,
pp.\ 3930--3945.

\bibitem{TD06}
Tian,~C.; Diggavi,~S.~N.~``Multistage successive refinement for Wyner-Ziv 
source coding with degraded side informations,'' 
{\em Proc.\ 2006 IEEE International Symposium on Information Theory (ISIT
2006)}, Seattle, WA, USA, 2006, 
pp. 1594-1598, doi: 10.1109/ISIT.2006.261545.

\bibitem{EGK11}
El Gamal, A.; Kim, Y.-H.~{\em Network Information Theory}, Cambridge
University Press, Cambridge, UK, 2011.

\bibitem{LZ76}
Lempel,~A.;~Ziv,~J.~``On the complexity of finite sequences,'' 
{\em IEEE Trans.~Inform.~Theory\/}, {\bf 1976}
vol.~IT--22, no.~1, pp.~137--143.

\bibitem{ZL78}
Ziv,~J.; Lempel,~A.~``Compression of individual sequences via 
variable-rate coding,''
{\em IEEE Trans.~Inform.~Theory\/} {\bf 1978},
vol.~IT--24, no.~5, pp.~530--536.

\bibitem{Ziv80}
Ziv,J.~``Distortion-rate theory for individual sequences,'' 
{\em IEEE Trans.~Inform.~Theory\/}, {\bf 1980}
vol.~IT--26, no.~2, pp.~137--143.

\bibitem{Ziv84}
Ziv,J.~``Fixed-rate encoding of individual sequences with side information,''
{\em IEEE Trans.~Inform.~Theory\/}, {\bf 1984},
vol.\ IT-30, no.\ 2, pp.\ 348--352.

\bibitem{YK96}
Yang,~E.-h.;~Kieffer,~J.~C.~``Simple universal lossy data compression schemes
derived from the Lempel-Ziv algorithm,'' {\it IEEE Trans.\ Inform.\ Theory}
{\bf 1996}, vol.\ 42, no.\ 1, pp.\ 239--245.

\bibitem{UK03}
Uyematsu,~T.; Kuzuoka, S,~``Conditional Lempel-Ziv 
complexity and its
application to source coding theorem with side information,''
{\em IEICE Trans.\ Fundamentals} {\bf 2003}, Vol.\ E86-A, no.\ 10, pp.\ 2615--2617.

\bibitem{me00}
Merhav,~N.~``Universal detection of messages via finite--state channels,''
{\it IEEE Trans.\ Inform.\ Theory} {\bf 2000},
vol.\ 46, no.\ 6, pp.\ 2242--2246.

\bibitem{p233}
Merhav,~N.~``A universal random coding ensemble for sample-wise lossy
compression,'' {\em Entropy}, 2023, 25(8), 1199;
{\tt https://doi.org/10.3390/e25081199},
August 2023.

\bibitem{p238}
Merhav,~N.~``Lossy compression of individual sequences revisited: fundamental
limits of finite-state encoders,'' {\em Entropy} {\bf 2024},
26(2), 116.\\
{\tt https://doi.org/10.3390/e26020116} January 28, 2024.

\bibitem{mesw}
Merhav,~N.~``Universal Slepian-Wolf coding for individual sequences,''
{\it IEEE Trans.\ Inform.\ Theory} {\bf 2025},
vol.\ 71, no.\ 1, pp.\ 783--796.

\bibitem{EGC82}
El Gamal, A.; Cover, T.~M.~``Achievable rates for multiple descriptions,''
{\it IEEE Trans.\ Inform.\ Theory}, {\bf 1982}, vol.\ 28, no.\ 6,
pp.\ 851--857.

\bibitem{ZB87}
Zhang, Z.; Berger, T.~``New results in binary multiple descriptions,''
{\it IEEE Trans.\ Inform.\ Theory}, {\bf 1987}, vol.\ 33, no.\ 4,
pp.\ 502--521.

\bibitem{me23}
Merhav, N.~``A universal random coding ensemble for sample-wise lossy
compression,'' {\em Entropy} 2023, 25, 1199.

\bibitem{Ziv85}
Ziv,~J.~ ``Universal decoding for finite-state channels,'' 
{\em IEEE Trans.~Inform.~Theory\/}, {\bf 1985},
vol.~IT--31, no.~4, pp.~453--460. 

\end{thebibliography}
\end{document}